\documentclass{article}
\usepackage{template/iclr2027/iclr2027_conference,times}

\usepackage{amsmath,amsfonts,bm}

\def\eqref#1{equation~\ref{#1}}

\def\1{\bm{1}}

\DeclareMathAlphabet{\mathsfit}{\encodingdefault}{\sfdefault}{m}{sl}
\SetMathAlphabet{\mathsfit}{bold}{\encodingdefault}{\sfdefault}{bx}{n}

\usepackage{amsmath,amssymb}
\usepackage{array}
\usepackage{booktabs}
\usepackage{graphicx}
\usepackage{subcaption}
\usepackage{float}
\usepackage{tabularx}
\usepackage{xcolor}
\usepackage{fontawesome5}
\usepackage{hyperref}
\usepackage{url}

\hypersetup{colorlinks=true,linkcolor=blue,citecolor=blue,urlcolor=blue}
\newcolumntype{Y}{>{\raggedright\arraybackslash}X}
\newcommand{\rooflang}{\textsc{RoofLang}}

\newcommand{\code}[1]{\texttt{#1}}

\title{RoofLang: Enabling AI-Driven Architecting of LLM Inference Systems}

\iclrfinalcopy
\author{
Ziyue Yang\textsuperscript{1}\thanks{Correspondence to: Ziyue Yang \texttt{<yzylivezh@hotmail.com>}.} \quad
Yuting Jiang\textsuperscript{1} \quad
Lei Qu\textsuperscript{1} \quad
Peng Cheng\textsuperscript{2}\\
\textsuperscript{1}Shanghai Xingyunzhili Artificial Intelligence Institute\\
\textsuperscript{2}Microsoft Research
}

\begin{document}
\maketitle
\fancyhead{}
\renewcommand{\headrulewidth}{0pt}

\begin{abstract}
AI is beginning to make substantive contributions to LLM inference
optimization. Existing AI optimizations are predominantly profiling-based.
Profiling-bound feedback confines the search to the capabilities and performance
of an existing software stack, preventing a fundamentally better architecture
of LLM inference systems from being identified. To enable the AI-driven LLM
inference system architecting loop, we argue that a general workload
representation, a verifiable mutation space, and an implementation-independent
evaluator are required. We present the \rooflang{}
domain-specific language (DSL) that provides these features. In our evaluation,
\rooflang{} reveals that DeepSeek V4-series models could achieve
3.5--39.5$\times$ higher peak decode throughput than other representative models.
This gap is disproportionate to their total parameter counts and arises largely
from compact KV-cache designs that support larger batches and reduce memory
traffic. A persistent optimizer agent further discovered several new
architectures that improved both throughput and interactivity of DeepSeek V4
Pro on NVIDIA B300 by 6.23--50.1\%.
\end{abstract}

\begin{center}
    \begin{tabular}{cl}
    \faGithub & \url{https://github.com/yzygitzh/rooflang} \\
     \faGlobe & \url{https://yzygitzh.github.io/rooflang}
    \end{tabular}
\end{center}

\section{Introduction}

AI systems are beginning to take on substantive parts of large language model
(LLM) inference optimization. Recent agents and automated workflows optimize
kernels, compiler-generated code, framework configurations, and serving paths,
turning implementation-level performance engineering into a machine-actionable
task \citep{ouyang2025kernelbench,flashinferbench,gpuscientist,
gai2026kernelpro,hanlab2026kda,inferencebench}. This emerging capability is an
important component of recursive self-improvement (RSI) of LLMs
\citep{schmidhuber2007godel,zhang2026dgm}.

Most existing approaches organize this work around profiling. Profiling
identifies hotspots in a current implementation and provides a clear verifier
for whether a local change helps
\citep{gpuscientist,gai2026kernelpro,hanlab2026kda,unger2022unity}. This feedback loop is
productive, but its potential is constrained by the software stack being
profiled. A fundamentally
different architecture may better satisfy the serving objective yet remain
unrealizable with the current stack because required features are absent or
their implementations are insufficiently optimized.

It would be desirable for AI to architect LLM inference systems from scratch and
continuously optimize them. Closing such an autonomous loop requires a general
representation of LLM workloads, a verifiable action space for architecting, and
a mechanism to evaluate resulting architectures and provide feedback. These
actions can organize computation and data through graph transformations,
sharding, placement, communication, and memory choices for a target workload
and hardware platform, potentially yielding more fundamental gains than tuning
a fixed implementation surface.

\begin{figure}[t]
\centering
\includegraphics[width=\linewidth]{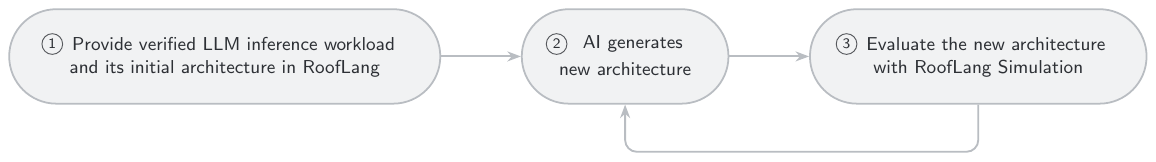}
\caption{AI-driven autonomous architecting loop enabled by RoofLang.}
\label{fig:autonomous-architecting-loop}
\end{figure}

To realize this vision, we present the \rooflang{} domain-specific language
(DSL), which provides these features. RoofLang represents workloads
and hardware as graphs, provides semantics-preserving graph transformations and
placement primitives as architecting actions, and evaluates the resulting
organizations with a roofline-based discrete-event simulator. The evaluator
exposes compute, memory, network, dependency, contention, lifetime, and capacity
costs while reducing interference from software-stack implementation details.
Given a verified initial workload and architecture, an AI agent can propose new
valid architecting actions and iteratively improve them using simulation
feedback, closing the autonomous loop to explore meaningful architecting
choices, as demonstrated in Figure~\ref{fig:autonomous-architecting-loop}.

Our evaluation reveals two salient findings. First, DeepSeek V4-series
models could achieve 3.5--39.5$\times$ higher peak decode throughput than other
representative models. Total parameter counts alone cannot explain this gap.
Instead, the advantage arises largely from compact KV-cache designs that
support larger batches and reduce memory traffic. Second, a persistent
optimizer agent discovered several new architectures that improved both
throughput and interactivity of DeepSeek V4 Pro on NVIDIA B300 by
6.23--50.1\%. Together, these results expose model-level design insights and
show that RoofLang can guide AI toward useful system architectures.

Our main contributions are as follows:
\begin{itemize}
\item \textbf{A practical DSL for autonomous LLM inference system
architecting.} RoofLang combines general graph representations, a
semantics-preserving action space, and a roofline-based evaluator in an
AI-accessible loop.
\item \textbf{An analysis of state-of-the-art models and architecting choices.}
We study how model, workload, and hardware characteristics affect graph
organization, placement, sharding, memory, communication, and their trade-offs.
\item \textbf{An evaluation of autonomous architecting by AI.} We compare
AI-driven exploration with heuristic grid search and examine how an agent uses
simulation feedback to generate meaningful architecting choices.
\end{itemize}

\section{RoofLang Design}

\subsection{Overview}

Architecting an LLM inference system can be viewed as organizing computation
and data to optimize a target objective. Autonomous architecting can be
realized with the following three elements:

\begin{itemize}
\item \textbf{A general workload representation.} A format sufficiently
general to describe LLM inference workloads without committing to a specific
software stack.
\item \textbf{Verifiable architecting actions.} This capability is necessary
to ensure that architecting actions made by autonomous processes are valid.
\item \textbf{An architecture evaluator.} A mechanism for evaluating
the quality of an architecting action while minimizing interference from
factors unrelated to architecting, such as software-stack implementation
details.
\end{itemize}

Given these elements and a verified input workload, an AI agent can generate
valid architecting actions, whose evaluation result then provides feedback for
revising those actions, enabling an autonomous loop for AI-driven architecting.

Compute graphs provide a general workload representation, as widely used in
modern AI frameworks \citep{abadi2016tensorflow}. Verified graph optimizers such
as Unity demonstrate how graph substitutions can preserve semantics while
changing computation and parallelization \citep{unger2022unity}. Our approach
combines graph-based workload and hardware representations, constrained
transformation and placement primitives, and roofline-based simulation
\citep{williams2009roofline}. This combination evaluates architecting choices
from analytical workload attributes and hardware specifications without
requiring profiled operator implementations.

\begin{figure}[H]
\centering
\includegraphics[width=\linewidth]{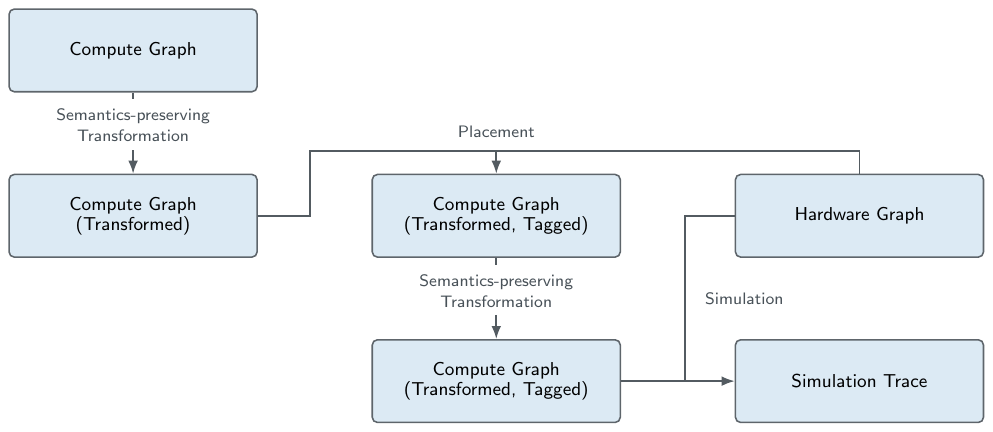}
\caption{RoofLang DSL overview.}
\label{fig:rooflang-framework}
\end{figure}

We therefore present the RoofLang DSL, which provides these features for
AI-driven architecting of LLM inference systems.
Figure~\ref{fig:rooflang-framework} gives an overview. A
correct and externally supplied compute graph is first rewritten through
semantics-preserving transformations. Placement combines the transformed
compute graph with a hardware graph by assigning operators and tensors to
hardware resources. Further semantics-preserving transformations can refine
the transformed and tagged compute graph, including its communication
realization. Finally, the transformed and tagged compute graph and the hardware
graph are evaluated by roofline-based simulation, which emits a trace as
feedback for subsequent architecting actions.

The following subsections introduce the different components of the RoofLang
DSL.

\subsection{Compute Graph}

The compute graph is an input to RoofLang and provides a way to
generally describe an LLM inference workload. It must be supplied externally, and
its provider is responsible for guaranteeing that it faithfully and correctly
represents the workload.

Each node in the compute graph is a compute node that represents an operation,
e.g., attention, GEMM, etc.
A node specifies its input and output tensors and has analytical workload
attributes such as tensor shapes and dtypes, FLOPs, and data movement. Compute
nodes are classified as side-effect-free or side-effecting: a side-effect-free node is
observable only through its output tensors, whereas a side-effecting node may
also modify externally observable state. Each edge in the compute graph
represents a data dependency from an output tensor of one operation to an input
tensor of another.

\subsection{Hardware Graph}

Another input to RoofLang is the hardware graph, which provides a
way to generally describe the target hardware platform for LLM inference. It
specifies the available compute and memory components and the fabrics that
connect them. The graph is supplied externally, and its provider is
responsible for guaranteeing that it faithfully and correctly represents the
target hardware.

Each node in the hardware graph is either a compute device (e.g., CPUs,
accelerators, NICs, switches, etc.) that has peak FLOPs for each supported
precision or a memory device (e.g., DRAM, HBM, SSDs, etc.) that has a specified
capacity. Each edge in the hardware graph represents a fabric connection
(e.g., PCIe, Ethernet, etc.) and has communication properties, including
bandwidth and latency.

\subsection{Placement}

Placement lowers a compute graph onto a hardware graph, tags operators with
compute devices and tensors with memory devices, so that it can be evaluated by
the simulator. This is realized by two
placement primitives summarized in Table~\ref{tab:placement-primitives}.
Note that placement is also semantics-preserving.

\begin{table}[t]
\caption{Placement primitives.}
\label{tab:placement-primitives}
\centering\small
\begin{tabularx}{\linewidth}{p{0.40\linewidth}Y}
\toprule
\textbf{Primitive} & \textbf{Function}\\
\midrule
\path{operator.set_compute} & Assigns an operator to the compute device and, optionally, the underlying stream/queue on which it executes.\\
\addlinespace[2pt]
\path{tensor.set_memory} & Assigns a tensor to the memory device in which it is stored.\\
\bottomrule
\end{tabularx}
\end{table}

\subsection{Semantics-preserving Transformation}

RoofLang implements semantics-preserving transformations through a set of
primitives over the compute graph, summarized in
Table~\ref{tab:graph-primitives}. Each primitive applies a pre-defined,
structurally constrained rewrite whose stated preconditions preserve the
modeled observations.

\begin{table}[t]
\caption{Semantics-preserving transformation primitives.}
\label{tab:graph-primitives}
\centering\small
\begin{tabularx}{\linewidth}{p{0.40\linewidth}Y}
\toprule
\textbf{Primitive} & \textbf{Function}\\
\midrule
\path{graph.add_control_edge} & Adds a happens-before dependency between two kernels.\\
\addlinespace[2pt]
\path{graph.remove_control_edge} & Removes a control edge between two kernels.\\
\midrule
\path{graph.insert_identity} & Insert an identity operator between two kernels.\\
\addlinespace[2pt]
\path{graph.remove_identity} & Remove an identity operator between two kernels.\\
\midrule
\path{graph.fuse_kernels} & Replaces a subgraph with an operator denoting the same composition at the external boundary, using pre-defined rules, e.g., qk-norm-rope fusion.\\
\addlinespace[2pt]
\begin{minipage}[t]{\linewidth}
\vspace{0pt}
\path{graph.split_kernel}
\end{minipage} &
\begin{minipage}[t]{\linewidth}
\vspace{0pt}
Replaces an operator with an equivalent partitioned or replicated subgraph
and proper communication operators, using pre-defined rules. Examples for
splitting a $(m,k)\times(k,n)$ GEMM are:
\begin{list}{\textbullet}{%
\setlength{\leftmargin}{1em}%
\setlength{\labelwidth}{0.7em}%
\setlength{\labelsep}{0.3em}%
\setlength{\itemsep}{0pt}%
\setlength{\parsep}{0pt}%
\setlength{\topsep}{0pt}%
}
\item split-by-$m$: scatter--GEMMs--gather
\item split-by-$k$: scatter--GEMMs--reduce
\item split-by-$n$: broadcast--GEMMs--gather
\end{list}
\end{minipage}\\
\midrule
\path{graph.dup_kernel} & Replicates a side-effect-free operator whose outputs are broadcasted, i.e., operator-broadcast $\Rightarrow$ broadcast-operators.\\
\addlinespace[2pt]
\path{graph.dedup_kernel} & Merge multiple same side-effect-free operators which accept broadcasted inputs, i.e., broadcast-operators $\Rightarrow$ operator-broadcast.\\
\midrule
\path{graph.merge_gather_scatter} & Replaces a gather-scatter subgraph with per-operator-pair direct data edges if splitting dimensions of both operators are the same, else with an all-to-all operator.\\
\addlinespace[2pt]
\path{graph.merge_gather_broadcast} & Replaces a gather-broadcast subgraph with an all-gather operator.\\
\addlinespace[2pt]
\path{graph.merge_reduce_scatter} & Replaces a reduce-scatter subgraph with a reduce-scatter operator.\\
\addlinespace[2pt]
\path{graph.merge_reduce_broadcast} & Replaces a reduce-broadcast subgraph with an all-reduce operator.\\
\bottomrule
\end{tabularx}
\end{table}

Executing a transformation updates the analytical attributes of the affected
operators and tensors. For example, operator FLOPs and bytes transferred, as
well as tensor shapes and communication volumes, are recomputed for the
transformed subgraph. Placement and semantics-preserving transformation can be
composed to express common LLM inference architecting actions. For example,
RoofLang can encode the following actions, with the relevant primitives
shown explicitly:

\begin{itemize}
\item \textbf{Prefill--decode (PD) disaggregation}: the primitives
\code{operator.set\_compute} and \code{tensor.set\_memory} assign prefill and
decode computation, together with exchanged state, to distinct compute and
memory devices \citep{zhong2024distserve}.
\item \textbf{KV-cache lifecycle}: \code{graph.dup\_kernel} and
\code{graph.dedup\_kernel} can be used to control the granularity of KV
recomputation and
caching, while identity operators such as \code{ToNvme}, \code{ToDram}, and
\code{ToHbm} can be inserted or removed with \code{graph.insert\_identity} and
\code{graph.remove\_identity} to describe different KV-cache memory-tier
policies \citep{sheng2023flexgen,kwon2023vllm}.
\item \textbf{Sharding}: \code{graph.split\_kernel},
\code{operator.set\_compute}, \code{tensor.set\_memory}, and communication
merging primitives can be used to realize typical model sharding mechanisms,
including tensor parallelism (TP), context parallelism (CP), data parallelism
(DP), expert parallelism (EP), and pipeline parallelism (PP)
\citep{narayanan2021megatron,jacobs2023ulysses,lepikhin2020gshard}.
\end{itemize}

\subsection{Simulation}
\label{sec:rooflang-simulation}

The simulator executes the transformed and tagged compute graph as a
topology-aware discrete-event simulation (DES). It obtains a topological order,
releases a kernel when all data and control predecessors have completed and
its assigned stream is available, and schedules start and completion events on
a priority queue. Each completion updates successor readiness, tensor
liveness, and resource occupancy, so the trace represents dependency-
constrained execution rather than a static sum of operator costs.

Memory footprints of inputs, outputs, and weights are tracked with operator
entry and exit events. The simulator reports the peak live bytes for each
memory device and raises an out-of-memory diagnostic when a capacity is
exceeded.

For an operator $K$ placed on compute device $d$, RoofLang models compute,
local-memory, and networking costs, respectively.

For compute cost, with $q$ indexing dtypes, $F_{K,q}$ denoting FLOPs, and
$P_{d,q}$ denoting the device peak FLOP/s at dtype $q$,
\begin{equation}
  T_{\mathrm{compute}}(K,d)
  = \sum_q \frac{F_{K,q}}{P_{d,q}}.
\end{equation}

For local-memory cost, including tensor reads from and writes to local memory,
\begin{equation}
  T_{\mathrm{memory}}(K,d)
  = \sum_{t\in\mathcal{R}_K}\frac{B^{\mathrm{read}}_{K,t}}
  {\beta^{\mathrm{read}}_{d,t}}
  + \sum_{t\in\mathcal{W}_K}\frac{B^{\mathrm{write}}_{K,t}}
  {\beta^{\mathrm{write}}_{d,t}},
\end{equation}
where the byte terms include the modeled input, weight, and output read/write
fractions.

For networking cost, including remote accesses and explicit communication,
data are routed through the directed hardware fabrics. The corresponding
network cost is
\begin{equation}
  T_{\mathrm{network}}(K)
  = \alpha_K +
  \begin{cases}
  \displaystyle\max_{\ell\in\mathcal{L}_K}
  \frac{B_{K,\ell}}{\beta_\ell}, & \text{remote tensor access},\\[6pt]
  \displaystyle\frac{V_K}{B_{\mathrm{eff},K}}, & \text{collective communication},
  \end{cases}
\end{equation}
with latency $\alpha_K$, link bytes $B_{K,\ell}$, link bandwidth
$\beta_\ell$, and collective volume $V_K$ and effective participant bandwidth
$B_{\mathrm{eff},K}$. In the no-contention baseline, the operator duration is
the bottleneck component, assuming ideal full overlap among compute,
local-memory, and networking costs (with byte and rate units chosen
consistently),
\begin{equation}
  T_K = \max\left\{T_{\mathrm{compute}},
  T_{\mathrm{memory}},T_{\mathrm{network}}\right\}.
\end{equation}

Collective communication uses a theoretical $\alpha+\beta$ model rather than a
packet-level schedule \citep{hockney1994communication}. The $\alpha$ term is the
maximum path latency among the participating ranks, i.e., the weighted diameter
of their data propagation graph. The $\beta$ term is an aggregate bandwidth
estimated heuristically from the hardware topology. Its effective value is
further controlled by the fabric resource-sharing mechanism described below,
so concurrent transfers reduce the available bandwidth according to contention.

The DES models resource sharing while events are active. Operators consuming a
compute device or a fabric receive normalized, capacity-weighted shares: if
$c_i$ is an operator's resource cap and $A_r$ is the active set consuming
resource $r$, its share is $c_i/\max(1,\sum_{j\in A_r}c_j)$. When sharing
occurs among multiple operators, the simulated cost during the
sharing interval extends proportionately until the set of sharing operators
changes. This gives a reasonable estimation of scenarios where multiple
operators share a single resource.

\section{Evaluation}

\subsection{RoofLang Efficacy on Representative LLMs and Hardware}
\label{sec:rooflang-efficacy}

\subsubsection{Setup}

\paragraph{Models.}
We evaluate four representative models: DeepSeek V4 Flash, GLM-5.3,
DeepSeek V4 Pro, and Kimi K3
\citep{deepseekai2026v4,glm5team2026glm5,zai2026glm53,kimiteam2026k3}.
Tables~\ref{tab:model-programs} and~\ref{tab:model-simulation} summarize their
model structures and their numerical precision configurations, respectively.
For simplicity, we omit KV-cache appending and speculative-decoding structures
from all four models, as well as the multimodal components of Kimi K3.
\begin{table}[t]
\caption{Models used in RoofLang evaluation.}
\label{tab:model-programs}
\centering\scriptsize
\setlength{\tabcolsep}{3pt}
\begin{tabularx}{\linewidth}{p{0.16\linewidth}p{0.13\linewidth}p{0.07\linewidth}p{0.13\linewidth}Y}
\toprule
\textbf{Model} & \textbf{Parameters} & \textbf{Layers} & \textbf{Attention} & \textbf{FFN}\\
\midrule
DeepSeek V4 Flash & 284BA13B & 43 & CSA + HCA & 1 shared and top-6 of 256 routed experts\\
GLM-5.3 & 744BA40B & 78 & DSA + MLA & 3 dense layers, 1 shared and top-8 of 256 routed experts\\
DeepSeek V4 Pro & 1.6TA49B & 61 & CSA + HCA & 1 shared and top-6 of 384 routed experts\\
Kimi K3 & 2.8TA104B & 93 & KDA + MLA & 1 dense layer, 2 shared and top-16 of 896 routed experts\\
\bottomrule
\end{tabularx}
\end{table}

\begin{table}[t]
\caption{Model-specific numerical precision configuration in RoofLang evaluation.}
\label{tab:model-simulation}
\centering\scriptsize
\setlength{\tabcolsep}{3pt}
\begin{tabularx}{\linewidth}{p{0.22\linewidth}p{0.34\linewidth}p{0.25\linewidth}Y}
\toprule
\textbf{Model} & \textbf{KV} & \textbf{Attention} & \textbf{FFN}\\
\midrule
DeepSeek V4 Flash & FP8 main cache and FP4 index cache & FP8 & FP4\\
GLM-5.3 & FP8 main cache and FP8 index cache & BF16 & FP8\\
DeepSeek V4 Pro & FP8 main cache and FP4 index cache & FP8 & FP4\\
Kimi K3 & FP8 MLA cache and BF16 KDA states & FP8 MLA and BF16 KDA & FP4\\
\bottomrule
\end{tabularx}
\end{table}

\paragraph{Hardware.}
Table~\ref{tab:hardware-params} summarizes four representative hardware
platforms: NVIDIA H200, NVIDIA GH200, NVIDIA B300, and NVIDIA GB300. We assume
NVLink for scale-up networking and InfiniBand for scale-out networking. For
NVIDIA H200 and NVIDIA GH200, they cannot do native FP4, and we assume they can
emulate FP4 as fast as FP8. The modeled peak rates, memory capacities, and
interconnect parameters are derived from NVIDIA specifications
\citep{nvidiah200,nvidia2026gh200,nvidia2026hgx,nvidia2025gb300}.

\begin{table}[t]
\caption{Hardware platforms used in RoofLang evaluation.}
\label{tab:hardware-params}
\centering
\scriptsize
\setlength{\tabcolsep}{2pt}
\resizebox{\linewidth}{!}{%
\begin{tabular}{lrrrrrrrrr}
\toprule
\textbf{Platform} & \shortstack{\textbf{FP4}\\\textbf{TFLOP/s}} &
\shortstack{\textbf{FP8}\\\textbf{TFLOP/s}} &
\shortstack{\textbf{BF16}\\\textbf{TFLOP/s}} &
\shortstack{\textbf{TF32}\\\textbf{TFLOP/s}} &
\shortstack{\textbf{HBM}\\\textbf{GB}} &
\shortstack{\textbf{HBM}\\\textbf{BW GB/s}} &
\shortstack{\textbf{NVLink}\\\textbf{BW GB/s}} &
\shortstack{\textbf{NVLink}\\\textbf{max scope}} &
\shortstack{\textbf{InfiniBand}\\\textbf{BW GB/s}}\\
\midrule
NVIDIA H200  & 1,979  & 1,979 & 989   & 495   & 144 & 4,800 & 450 & 8   & 50  \\
NVIDIA GH200 & 1,979  & 1,979 & 989   & 495   & 144 & 4,900 & 450 & 256 & 50  \\
NVIDIA B300  & 13,500 & 4,500 & 2,250 & 1,125 & 288 & 7,750 & 900 & 8   & 100 \\
NVIDIA GB300 & 15,000 & 5,000 & 2,500 & 1,250 & 288 & 8,000 & 900 & 72  & 100 \\
\bottomrule
\end{tabular}
}
\end{table}

\paragraph{Workload.}
Table~\ref{tab:evaluation-matrix} summarizes the workload space evaluated,
including stages, contexts, GPU counts, parallelism dimensions, and batch
sizes. For GPU counts, we select 8 and 64 to
probe the efficacy of larger scale-up networking scope. For context length,
we select 64K as it is the dominant long-context regime according to a recent
study \cite{lee2026dcp}, and we select 1M as it is the maximum supported by the
evaluated models
\citep{deepseekai2026v4,zai2026glm53,kimiteam2026k3}. We adopt context
parallelism (CP) instead of tensor parallelism (TP) because sequence-wise
sharding is better suited to long contexts and avoids the KV-cache replication
that TP can induce \cite{lee2026dcp}. We further heuristically let
$\mathrm{DP}\times\mathrm{CP}=\mathrm{EP}$, referencing the large-scale inference
configuration in the DeepSeek-V3 report \cite{deepseekai2024v3}. For batch
size, we sweep $2^n$ from $n=0$ and stop at the first OOM point.

\begin{table}[t]
\caption{Workload configuration search space in RoofLang evaluation.}
\label{tab:evaluation-matrix}
\centering\small
\setlength{\tabcolsep}{4pt}
\renewcommand{\arraystretch}{0.95}
\begin{tabularx}{0.92\linewidth}{p{0.19\linewidth}Y}
\toprule
\textbf{Configuration} & \textbf{Values}\\
\midrule
Stage & prefill, decode (1-step)\\
Context & 64K, 1M\\
GPU counts & 8, 64\\
Parallelism & DP, CP, PP, EP, where
$\mathrm{EP}\times\mathrm{PP}=\mathrm{DP}\times\mathrm{CP}\times\mathrm{PP}=n_{\mathrm{GPUs}}$\\
Batch size & $2^n$ until OOM, $n\geq 0$\\
\bottomrule
\end{tabularx}
\end{table}

\paragraph{Simulation.}
We run the discrete-event simulation (DES) as described in
Section~\ref{sec:rooflang-simulation}. Across the overall execution timeline,
we assume that computation and communication overlap perfectly through
fine-grained overlapping \citep{deepseekai2024v3}, and derive the final
throughput and interactivity from the overlapped timeline. We further assume
that routed tokens are ideally balanced across MoE experts.

\subsubsection{Results}

\begin{figure*}[!t]
\centering
\begin{subfigure}[t]{0.49\linewidth}
\centering
\includegraphics[width=\linewidth]{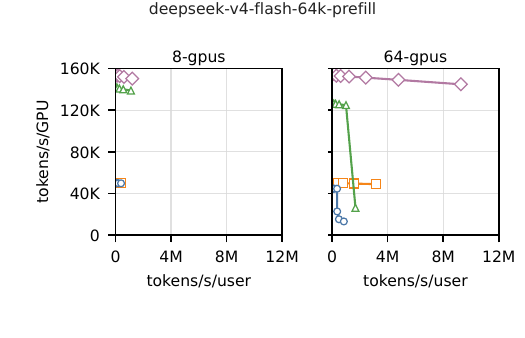}
\end{subfigure}\hfill
\begin{subfigure}[t]{0.49\linewidth}
\centering
\includegraphics[width=\linewidth]{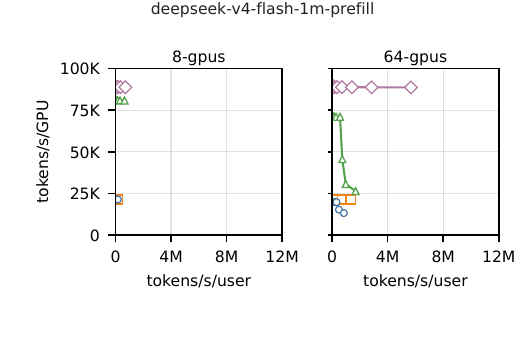}
\end{subfigure}
\par\vspace{0.25em}
\begin{subfigure}[t]{0.49\linewidth}
\centering
\includegraphics[width=\linewidth]{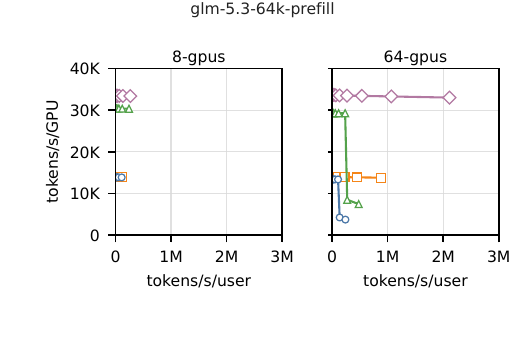}
\end{subfigure}\hfill
\begin{subfigure}[t]{0.49\linewidth}
\centering
\includegraphics[width=\linewidth]{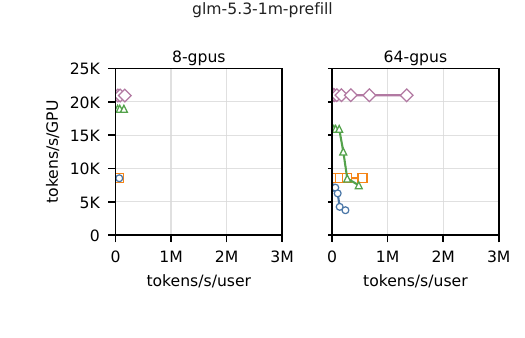}
\end{subfigure}
\par\vspace{0.25em}
\begin{subfigure}[t]{0.49\linewidth}
\centering
\includegraphics[width=\linewidth]{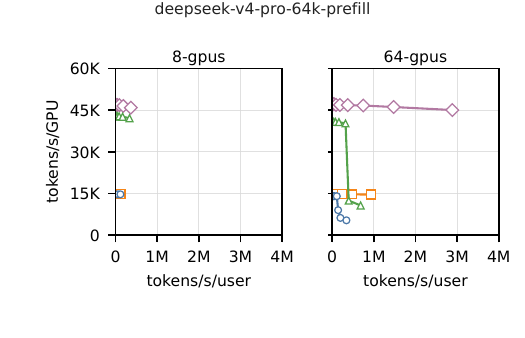}
\end{subfigure}\hfill
\begin{subfigure}[t]{0.49\linewidth}
\centering
\includegraphics[width=\linewidth]{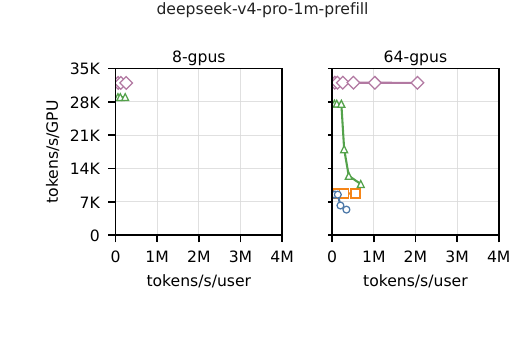}
\end{subfigure}
\par\vspace{0.25em}
\begin{subfigure}[t]{0.49\linewidth}
\centering
\includegraphics[width=\linewidth]{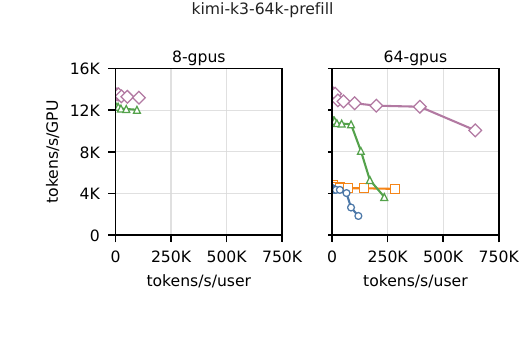}
\end{subfigure}\hfill
\begin{subfigure}[t]{0.49\linewidth}
\centering
\includegraphics[width=\linewidth]{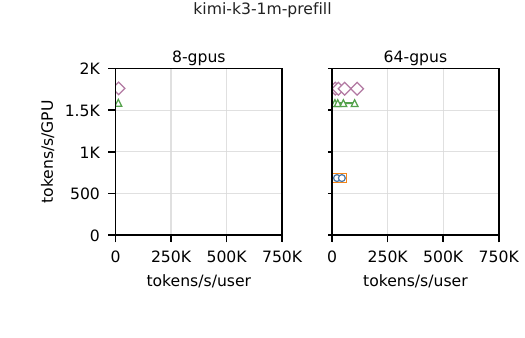}
\end{subfigure}
\par\vspace{0.25em}
\includegraphics[width=0.42\linewidth]{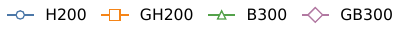}
\caption{Throughput (token/s/GPU) - interactivity (token/s/user) pareto frontier for prefill.}
\label{fig:pareto-prefill}
\end{figure*}

\begin{figure*}[!t]
\centering
\begin{subfigure}[t]{0.49\linewidth}
\centering
\includegraphics[width=\linewidth]{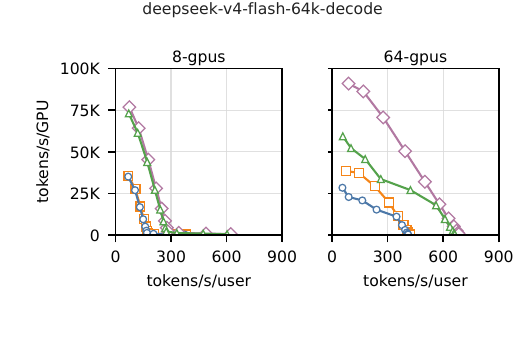}
\end{subfigure}\hfill
\begin{subfigure}[t]{0.49\linewidth}
\centering
\includegraphics[width=\linewidth]{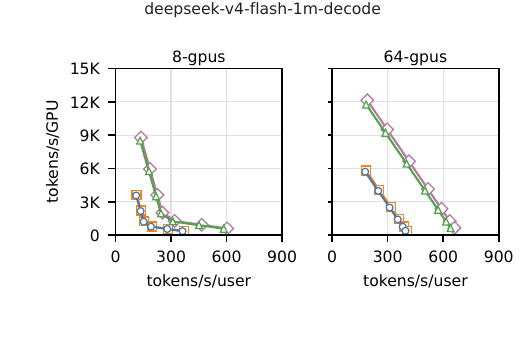}
\end{subfigure}
\par\vspace{0.25em}
\begin{subfigure}[t]{0.49\linewidth}
\centering
\includegraphics[width=\linewidth]{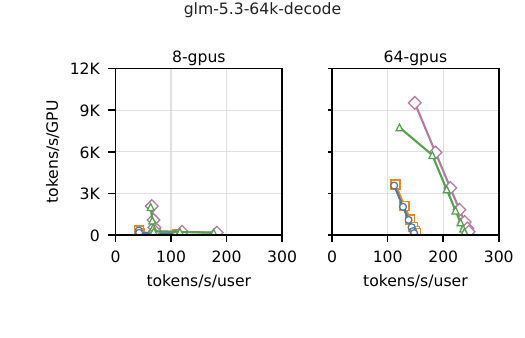}
\end{subfigure}\hfill
\begin{subfigure}[t]{0.49\linewidth}
\centering
\includegraphics[width=\linewidth]{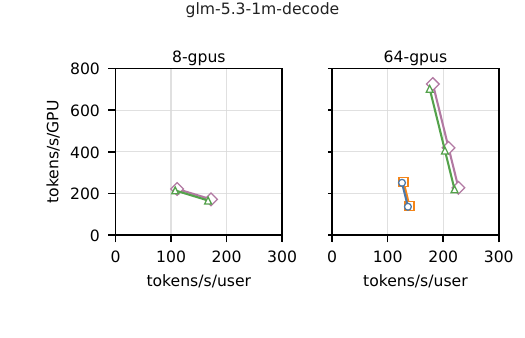}
\end{subfigure}
\par\vspace{0.25em}
\begin{subfigure}[t]{0.49\linewidth}
\centering
\includegraphics[width=\linewidth]{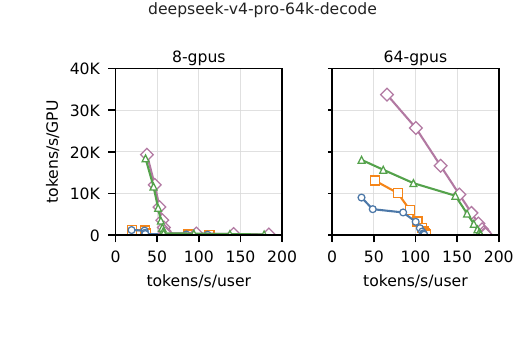}
\end{subfigure}\hfill
\begin{subfigure}[t]{0.49\linewidth}
\centering
\includegraphics[width=\linewidth]{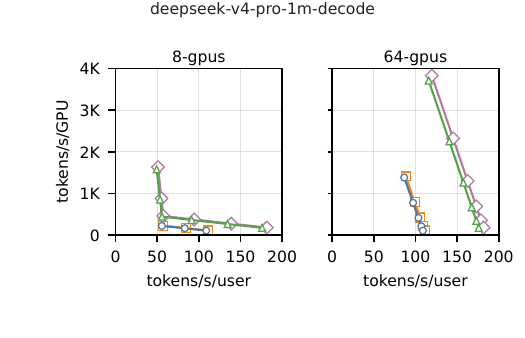}
\end{subfigure}
\par\vspace{0.25em}
\begin{subfigure}[t]{0.49\linewidth}
\centering
\includegraphics[width=\linewidth]{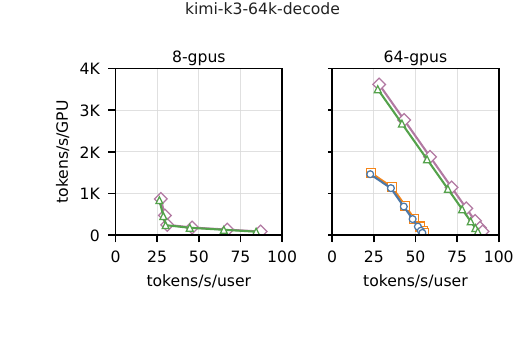}
\end{subfigure}\hfill
\begin{subfigure}[t]{0.49\linewidth}
\centering
\includegraphics[width=\linewidth]{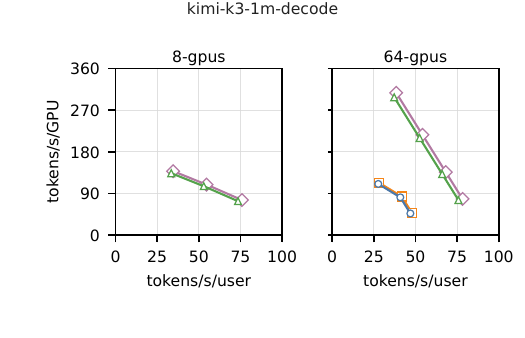}
\end{subfigure}
\par\vspace{0.25em}
\includegraphics[width=0.42\linewidth]{figures/bin/evaluation-hardware-legend.pdf}
\caption{Throughput (token/s/GPU) - interactivity (token/s/user) pareto frontier for decode.}
\label{fig:pareto-decode}
\end{figure*}

Figures~\ref{fig:pareto-prefill} and~\ref{fig:pareto-decode} plot the resulting
throughput--interactivity Pareto frontiers for prefill and decode, respectively.

\paragraph{Frontier shape.}
Across models, contexts, hardware platforms, and GPU counts, the curves exhibit
the expected throughput--interactivity shape: higher throughput coincides with
lower interactivity, and vice versa.

\paragraph{Frontier endpoints.}
The left endpoints for prefill correspond to large-batch,
high-arithmetic-intensity configurations, and their throughput values are
primarily bounded by model compute FLOPs divided by GPU FLOP rates. The right
endpoints of decode correspond to small-batch, interactivity-oriented
configurations, and their interactivity values are primarily bounded by the
activated-parameter bytes per step divided by HBM bandwidth.

\paragraph{Decode throughput.}
Among the plotted curves for 64 NVIDIA GB300 GPUs, at 64K-context and
1M-context, respectively, DeepSeek V4 Flash
reaches 9.6$\times$ and 16.7$\times$ the peak decode throughput of GLM-5.3.
The corresponding ratios over Kimi K3 are 25.2$\times$ and 39.5$\times$. For
DeepSeek V4 Pro, the 64K and 1M ratios are 3.5$\times$ and 5.3$\times$ over
GLM-5.3, and 9.3$\times$ and 12.4$\times$ over Kimi K3. These gaps arise in
large part from per-decode-token memory overhead. For batch size $B$, we
calculate this overhead as
\begin{equation}
M_{\mathrm{decode}} = \frac{P + B T_{\mathrm{KV}}}{B},
\end{equation}
where $P$ is the encoded model-weight size and $T_{\mathrm{KV}}$ is the
per-request total KV-cache read. The peak decode throughput in
Figure~\ref{fig:pareto-decode} is thus largely determined by the
per-decode-token memory traffic in Tables~\ref{tab:kv-cache-sizes-64k}
and~\ref{tab:kv-cache-sizes-1m}. Models with
smaller per-request KV-cache footprints can sustain larger batches, more
effectively amortizing model-weight traffic in the memory-bound regime.
Smaller per-request KV-cache traffic further reduces $M_{\mathrm{decode}}$ and
amplifies this throughput advantage.

\paragraph{Interconnect sensitivity.}
Larger scale-up networking domains enable faster MoE token routing and are more
beneficial when MoE routing traffic dominates. This aligns with the expected
bottleneck shifting between attention and MoE. For prefill, computation dominates,
so MoE routing traffic can be readily hidden at reasonably large batch sizes
without substantially affecting interactivity. For decode, KV-cache traffic
under CP is carried by aggregation collectives and runs at similar bandwidth
over scale-up and scale-out networks. As the KV cache grows, MoE token routing
accounts for a smaller fraction of the total overhead because attention time
increases and the routing traffic itself decreases with the lower sustainable
batch size, leaving less end-to-end room for a larger scale-up networking domain
to improve performance. The decode curves in Figure~\ref{fig:pareto-decode} follow the
KV-cache traffic and batch sizes at peak throughput reported in
Tables~\ref{tab:kv-cache-sizes-64k} and~\ref{tab:kv-cache-sizes-1m}.

\begin{table}[t]
\caption{Per-decode-token memory overhead at 64K context
(64 NVIDIA GB300 GPUs).}
\label{tab:kv-cache-sizes-64k}
\centering
\scriptsize
\setlength{\tabcolsep}{3pt}
\begin{tabular*}{\linewidth}{@{\extracolsep{\fill}}lrrrrr}
\toprule
\textbf{Model}
& \shortstack{\textbf{Per-request}\\\textbf{parameters traffic}}
& \shortstack{\textbf{Per-request}\\\textbf{KV-cache footprint}}
& \shortstack{\textbf{Per-request}\\\textbf{KV-cache traffic}}
& \shortstack{\textbf{Peak}\\\textbf{batch size}}
& \shortstack{\textbf{Per-decode-token}\\\textbf{memory traffic}}\\
\midrule
DeepSeek V4 Flash & 2.277~MiB & 0.192~GiB & 0.033~GiB & 65,536 & 0.035~GiB\\
GLM-5.3 & 173.618~MiB & 2.906~GiB & 0.250~GiB & 4,096 & 0.419~GiB\\
DeepSeek V4 Pro & 24.811~MiB & 0.275~GiB & 0.055~GiB & 32,768 & 0.079~GiB\\
Kimi K3 & 175.897~MiB & 1.065~GiB & 1.065~GiB & 8,192 & 1.237~GiB\\
\bottomrule
\end{tabular*}
\end{table}

\begin{table}[t]
\caption{Per-decode-token memory overhead at 1M context
(64 NVIDIA GB300 GPUs).}
\label{tab:kv-cache-sizes-1m}
\centering
\scriptsize
\setlength{\tabcolsep}{3pt}
\begin{tabular*}{\linewidth}{@{\extracolsep{\fill}}lrrrrr}
\toprule
\textbf{Model}
& \shortstack{\textbf{Per-request}\\\textbf{parameters traffic}}
& \shortstack{\textbf{Per-request}\\\textbf{KV-cache footprint}}
& \shortstack{\textbf{Per-request}\\\textbf{KV-cache traffic}}
& \shortstack{\textbf{Peak}\\\textbf{batch size}}
& \shortstack{\textbf{Per-decode-token}\\\textbf{memory traffic}}\\
\midrule
DeepSeek V4 Flash & 36.439~MiB & 3.034~GiB & 0.414~GiB & 4,096 & 0.450~GiB\\
GLM-5.3 & 2,777.892~MiB & 46.500~GiB & 2.711~GiB & 256 & 5.423~GiB\\
DeepSeek V4 Pro & 396.984~MiB & 4.344~GiB & 0.608~GiB & 2,048 & 0.996~GiB\\
Kimi K3 & 2,814.346~MiB & 13.721~GiB & 13.721~GiB & 512 & 16.469~GiB\\
\bottomrule
\end{tabular*}
\end{table}

\subsection{AI-Driven Autonomous Architecting}

\subsubsection{Setup}

We implement a simple optimizer agent by running Claude Code persistently in
non-interactive mode, without a preset turn limit
\citep{anthropicclaudecodecli}, thereby realizing the autonomous loop in
Figure~\ref{fig:autonomous-architecting-loop}. Following DeepSeek's official
Claude Code integration guide, we configure DeepSeek V4 Pro as the primary
model and DeepSeek V4 Flash as the lightweight and subagent model
\citep{deepseekclaudecode}.

We let the optimizer agent improve the DeepSeek V4 Pro implementation in
RoofLang. In each optimization session, we provide the agent with the RoofLang
source repository, including the DeepSeek V4 Pro implementation, together with the
baseline results from Section~\ref{sec:rooflang-efficacy}. The agent is
instructed to keep seeking new device placements and semantics-preserving graph
transformations that improve the throughput--interactivity Pareto frontier
according to simulation feedback. It continues this loop until it decides to
stop.

\subsubsection{Results}

We discuss three representative cases below.

\paragraph{Cost- and head-aware contiguous PP partition.}
Balancing PP stages solely by layer count can leave the embedding-bearing first
stage or output-head-bearing last stage on the critical path. The agent
therefore calibrated stage costs from traces and enumerated legal contiguous
partitions that reduced either or both boundary-stage loads without changing
layer order or PP degree. For DeepSeek V4 Pro decode-64K on 256 B300 GPUs with
batch 4,096, EP=16, and PP=16, it replaced the balanced partition
$(4\times13,3\times3)$ with $(3,4\times14,2)$. Accounting for the extra
embedding and output-head work reduced the maximum stage span and improved both
throughput and interactivity by 20.3\%.

\paragraph{DP--CP routed-expert splitting.}
The agent split each routed expert along the existing DP-major, CP-minor token
shards and applied \code{graph.split\_kernel} to the expert up and down
projections, creating one copy per DP$\times$CP shard. This allowed
same-dimension Gather--Scatter pairs to be replaced by rank-local data edges,
removing redundant local token materialization while preserving routing and
expert ownership. Because the split incurs additional expert-weight reads, it
was enabled only when each copy had at least 1,024 routed tokens and EP was at
most eight. For DeepSeek V4 Pro prefill-8K on eight B300 GPUs with batch 512,
the large prefill token volume amortized the extra weight traffic, and removing
the local Gather--Scatter traffic improved both throughput and interactivity by
6.23\%.

\paragraph{Node-local expert replication.}
The agent replicated selected experts across the two nodes of an EP=16 stage,
so routed-token dispatch and combine paths used node-local NVLink rather than
cross-node InfiniBand. The change preserves expert computation but trades
additional expert-weight reads and HBM capacity for lower exposed routing
communication, and was applied only to communication-bound cases. For
DeepSeek V4 Pro prefill-1M on 16 B300 GPUs with batch 1 and CP=EP=16, the large
routed-token shards made inter-node communication dominant; node-local routing
reduced that bottleneck and improved both throughput and interactivity by
50.1\%.

\section{Related Work}

\subsection{Inference Performance Modeling}

Inference performance models span analytical abstractions and high-fidelity
simulators. The Roofline model bounds performance using arithmetic intensity
and hardware compute and bandwidth ceilings \citep{williams2009roofline}, while
LLM-Viewer and Calculon apply analytical modeling to LLM inference
\citep{yuan2024llmviewer,isaev2023calculon}. Simulation systems incorporate more
execution detail: ASTRA-sim models distributed computation and communication,
Vidur simulates LLM serving workloads, Frontier uses discrete-event
simulation with timing and memory models calibrated against concrete serving
stacks, and KernelSight-LM
augments a kernel-level roofline model with learned efficiency factors and
embeds it in a discrete-event serving simulator
\citep{won2023astrasim,agrawal2024vidur,feng2026frontier,yao2026kernelsightlm}.
These methods trade evaluation cost and generality for different levels of fidelity, but
performance estimation alone does not establish semantic equivalence between
candidate architectures.

Verified graph optimization addresses this complementary requirement. Unity
jointly optimizes algebraic transformations and parallelization for distributed
DNN training using a unified parallel computation graph \citep{unger2022unity}.

\rooflang{} combines a constrained mutation space with a roofline-based
evaluator derived from analytical workload attributes and hardware
specifications, without requiring profiled operator implementations. Its focus
is this combination for AI-driven LLM inference system architecting, rather
than transformation verification alone. Human verification remains necessary
for the initial input graphs and the transformation contracts; subsequent
exploration is restricted to placement operations and semantics-preserving
transformations under their stated preconditions.

\subsection{Kernel Agents and Other Efforts for RSI}

Kernel agents and related efforts toward recursive self-improvement (RSI)
close optimization loops around executable implementations
\citep{schmidhuber2007godel,zhang2026dgm}. SGLang exposes profiling tools and
agent-facing developer skills \citep{sglangprofiling,sglangagentskills}, while
Kernel Design Agents and KDA-Pilot organize repository-scale kernel optimization
around task contracts, correctness gates, profiling evidence, and repeated
benchmarking
\citep{hanlab2026kda,bbuf2026kdapilot}. KernelBench and FlashInfer-Bench
evaluate generated kernels through execution and performance tests
\citep{ouyang2025kernelbench,flashinferbench}, while GPU Kernel Scientist and
KernelPro iteratively optimize kernels using profiling feedback
\citep{gpuscientist,gai2026kernelpro}. AMD GEAK provides a related industrial
agentic workflow \citep{amdgeak}. InferenceBench extends this approach to
open-ended inference-server optimization \citep{inferencebench}. OptiFlow
optimizes collective communication by combining LLM-generated data-movement
programs in a DSL, deterministic scheduling, and iterative refinement using
real-hardware feedback \citep{long2026optiflow}. These systems
either manipulate real code directly or derive feedback from performance
samples collected on real hardware, rather than using an
implementation-independent architectural oracle.

This implementation-grounded feedback is concrete, but it couples the search
space to the available kernels, compiler, framework, and hardware. A design may
remain unreachable or appear inferior when its required features are absent or
poorly optimized in the current stack. \rooflang{} decouples architectural
design from concrete implementation details by evaluating graph organizations
with hardware specifications and roofline relations. It can therefore search
for designs that are intrinsically more efficient under theoretical resource
limits before implementation quality biases the comparison. After an
architecture is selected, kernel agents and framework-level tools can optimize
its concrete implementation.

\section{Conclusion, Limitations, and Future Directions}

\rooflang{} is an initial step toward enabling AI-driven architecting of LLM
inference systems. The DSL combines
expert-authored compute and hardware graphs, explicit placement and
communication, semantics-preserving transformations, and a roofline-based
discrete-event simulator. Its evaluation yields two central insights.
First, DeepSeek V4-series models could achieve 3.5--39.5$\times$ higher peak decode
throughput than other representative models, despite no commensurate difference
in total parameter counts. This advantage arises largely from compact KV-cache
designs that support larger batches and reduce memory traffic. Second, a
persistent optimizer agent discovered several new architectures that improved
both throughput and interactivity of DeepSeek V4 Pro on NVIDIA B300 by
6.23--50.1\%. Together, these findings indicate that a verification-constrained
analytical model can expose useful model-level insights and guide architectural
exploration before implementation-specific optimization.

Despite this progress, three limitations currently bound the autonomy,
modeling fidelity, and scalability of \rooflang{}.

\paragraph{Initial-input verification.}
\rooflang{} currently assumes that the initial compute and hardware graphs are
correct. This concentrates but does not eliminate the need for human
verification. Future work should automate initial-input construction and
validation, for example by deriving compute graphs from official baseline
implementations and checking them against execution traces.

\paragraph{Modeling fidelity.}
The current models simplify behaviors that can materially affect performance.
They assume ideal MoE load balance and omit KV-cache appending,
speculative-decoding structures, DeepSeek V4's mHC, and Kimi K3's multimodal
components. Future extensions should represent these behaviors explicitly
and calibrate them against measured implementations while preserving the
existing mutation and equivalence contracts.

\paragraph{Simulation scalability.}
The current discrete-event simulator does not scale efficiently to the largest
configurations. Simulating Kimi K3 on 128 GPUs can require hours and hundreds
of gigabytes of memory, which makes repeated agentic optimization impractical.
Future work should reduce event and state materialization, reuse unchanged
subgraphs across candidates, and introduce scalable parallel or approximate
evaluation without weakening correctness checks.

\bibliographystyle{template/iclr2027/iclr2027_conference}
\bibliography{references}

\end{document}